\documentclass[a4paper]{jpconf}
\usepackage{graphicx}
\begin{document}
\title{Electromagnetic Decay of Molecular States }

\author{S. Courtin, A. Goasduff and F. Haas}
\address{Institut Pluridisciplinaire Hubert Curien, UMR 7178, Universit\'e de Strasbourg / CNRS-IN2P3. 23, rue du Loess, 67037 Strasbourg, France}
\ead{sandrine.courtin@iphc.cnrs.fr}

\begin{abstract}
Electromagnetic transitions from deformed structures based on $\alpha$ configurations or on heavier clusters are discussed, drawing the link between multiparticle-multihole excited bands and cluster structures. Enhanced E2 and E1 transitions are reviewed in the light nuclei, $^8$Be, $^{10}$Be, $^{12}$C, $^{16}$O, $^{18}$O and heavier ones like $^{212}$Po. Connections between cluster structures and superdeformed configurations in $^{36}$Ar and $^{40}$Ca are discussed. What the cluster states based on heavier substructures like $^{12}$C and $^{16}$O are concerned, recent results on the resonant radiative capture reaction $^{12}$C($^{16}$O,$\gamma$)$^{28}$Si are presented, in particular the strong decay mode involving the feeding of low-lying $^{28}$Si 1$^+$ and 2$^+$ T=1 states by enhanced M1 isovector transitions.
\end{abstract}

\section{Introduction}
Throughout the nuclear chart, there exists well identified regions of nuclei with large deformations. For such nuclei, the characteristic signature of deformation is the grouping of excited states into rotational bands with large moment of inertia. The deformed feature of an excited band member generally manifests itself through collective, i.e. enhanced, E2 intraband transitions. In light nuclei, as we will show later, deformation is very often connected with clusterization. To assess such connection, it is thus highly desirable to search for the electromagnetic decay mode of cluster states, sometimes also called molecular states. The `basic' cluster states in light nuclei are composed of $\alpha$ particles and we will now briefly recall some `highlight' cases for which enhanced E2 and also sometimes E1 gamma transitions have been observed. We will then discuss the occurrence of heavier clusters in the sd shell, and in particular the $^{12}$C-$^{16}$O resonances in $^{28}$Si that we have explored via resonant radiative capture experiments.

\section{Alpha cluster bands probed by E1 and E2 transitions}
Among the light even-even N=Z nuclei, $^8$Be is certainly the `archetype' example of an $\alpha$-$\alpha$ nucleus showing a band structure composed of the 0$^+$ ground state and excited 2$^+$ and 4$^+$ states at 3.04 and 11.4 MeV, respectively. These three states mainly decay into 2$\alpha$ particles, but a strong 4$^+$ $\rightarrow$ 2$^+$ $\gamma$ transition has been observed with a reduced B(E2) strength of 25.8 $\pm$ 8.4 W.u. \cite{datar}. This very important result in the context of $\gamma$ decay of molecular states, has been obtained through the radiative capture reaction $\alpha$($\alpha$,$\gamma$) and an $\alpha$-$\alpha$-$\gamma$ coincidence selection in the exit channel \cite{datar}. The E2 2$^+$ $\rightarrow$ 0$^+$ $\gamma$ transition remains to be measured and is clearly a challenging goal.\\
In $^9$Be, there is a well developped $\alpha$-$\alpha$-n cluster band built on the J$^{\pi}$ = $\frac{3}{2}^-$ ground state. The excited band members with  J$^{\pi}$ = $\frac{5}{2}^-$ and  $\frac{7}{2}^-$ are located at E$_x$ = 2.43 and 6.38 MeV, respectively. The intra cluster band transitions $\frac{5}{2}^-$ $\rightarrow$ $\frac{3}{2}^-$ and $\frac{7}{2}^-$ $\rightarrow$ $\frac{5}{2}^-$ are highly collective with B(E2) = 24.4 $\pm$ 1.8 W.u. and 8.5 $\pm$ 3.6 W.u. \cite{nndc}, confirming thus the cluster, i.e. deformed, structure of this nucleus.\\
$^{10}$Be is among light nuclei one of the best example of clustering. However in this case, the $\alpha$-$\alpha$-n-n clustering is mainly concentrated in two excited bands, one with K$^{\pi}$ = 0$^+$ and J$^{\pi}$ = 0$^+$, 2$^+$ and 4$^+$ band members at 6.18, 7.54 and 10.15 MeV, respectively and another one with K$^{\pi}$ = 1$^-$ and  J$^{\pi}$ = 1$^-$, 2$^-$, 3$^-$ and 4$^-$ at E$_x$ = 5.96, 6.26, 7.37 and 9.27 MeV, respectively \cite{mil1,freer,mil2}. The first excited 0$^+$ state at 6.18 MeV which is the band-head of the K$^{\pi}$ = 0$^+$ $\alpha$-$\alpha$-n-n cluster band cannot be reproduced by conventional shell model calculations which always place the first 0$^+$ excited state of $^{10}$Be at excitation energies above 10 MeV. This cluster 0$^+$ state decays by a very moderate E2 transition of B(E2) = 2.6 $\pm$ 0.9 W.u. to the first $^{10}$Be 2$^+$ state. But it also decays to the cluster 1$^-$ state of the K$^{\pi}$ = 1$^-$ band by a $\gamma$ transition of E$_{\gamma}$ = 219 keV and a very large E1 strength of 4.3  $\pm$ 1.6 $\times$ 10$^{-2}$ W.u.
\cite{nndc}, indeed one of the strongest E1 transitions for nuclei in the p shell. We will come back to these strong E1 transitions between cluster states a little later. These two positive and negative parity $\alpha$-$\alpha$-n-n cluster bands have a much larger moment of inertia than the cluster bands in $^{8}$Be($\alpha$-$\alpha$) and $^{9}$Be($\alpha$-$\alpha$-n). In this case, the intraband E2 transitions should be searched for in the K$^{\pi}$ = 0$^+$ band between the 7.54 MeV, J$^{\pi}$ = 2$^+$ state and the 6.18 MeV,  J$^{\pi}$ = 0$^+$ state and also in the K$^{\pi}$ = 1$^-$ band between the 7.37 MeV,  J$^{\pi}$ = 3$^-$ state and the 5.96 MeV, J$^{\pi}$ = 1$^-$ state.\\
What $\alpha$ clustering is concerned, $^{12}$C is certainly a textbook example. The first excited 0$^+$ state is located just above the $^{8}$Be+$\alpha$ threshold, it has an excitation energy of 7.65 MeV and an almost pure $\alpha$-$\alpha$-$\alpha$ cluster configuration. This so-called `Hoyle' state decays mainly through the 3$\alpha$ channel but has also small decay branches to the 0$^+_{gs}$ (E0 decay) and to the first 2$^+$ state at 4.43 MeV (E2 decay). The E2 strength of 8.1 $\pm$ 1.1 W.u. \cite{nndc} corresponds to a moderately collective transition between the very deformed 0$^+$ $\alpha$-$\alpha$-$\alpha$ `Hoyle' state and the almost pure 2$^+$ shell model `p' state. Between the slightly oblate deformed $^{12}$C first 2$^+$ and ground state, we have B(E2) = 4.7 $\pm$ 0.2 W.u. \cite{nndc}. Possible excited states built on the `Hoyle' state are actually actively searched for and candidates for the 2$^+$ and 4$^+$ members of the `Hoyle' band are reported at excitation energies of $\sim$ 9.8 and $\sim$ 13.3 MeV, respectively \cite{freer2,ito}.  AMD model predictions of E2 transitions inside the `Hoyle' band have been performed by Kanada-En'Yo \cite{kan1} and values of 61.3 and 376.6 W.u. are obtained for the 2$^+$ $\rightarrow$ 0$^+$ and 4$^+$ $\rightarrow$ 2$^+$ E2 transitions, respectively. It turns out that it is very difficult to locate experimentally and characterize the excited 2$^+$ and 4$^+$ `Hoyle' states which are at high excitation energies and surrounded by several other states which generally have a large natural width. It will thus probably be experimentally difficult, if not impossible, to measure and eventually confirm these predicted very enhanced E2 transitions between these strongly deformed $\alpha$-$\alpha$-$\alpha$ cluster `Hoyle' states.\\
$^{16}$O is a doubly-magic nucleus with N=Z=8 and with a first excitd 0$^+$ state at 6.05 MeV. This state is the band head of a cluster $\alpha$-$^{12}$C band with the 2$^+$ and 4$^+$ band members at 6.92 and 10.36 MeV, respectively. These $\alpha$ cluster states are strongly excited in Li induced $\alpha$ transfer reactions on $^{12}$C \cite{nndc}, the unbound 4$^+$ state at 10.35 MeV has also been observed as a strong resonance in the $^{12}$C($\alpha$,$\gamma$) capture reaction. This $\alpha$-$^{12}$C cluster band shows remakably strong E2 strengths of 27 $\pm$ 3 W.u. for the 6.92(2$^+)$ $\rightarrow$ 0(0$^+$) transition and of 65 $\pm$ 6 W.u. for the 10.36(4$^+)$ $\rightarrow$ 6.92(2$^+$) transition \cite{nndc}. These E2 strengths are among the strongest in the whole sd shell. Generally cluster states cannot be described by shell model (SM) calculations. In the case of $^{16}$O, the SM calculations have been rather successful and the cluster $\alpha$ band has been clearly identified with a 4p-4h structure using the SM ZBM interaction with a $^{12}$C core \cite{zbm}. $^{16}$O was the first case where the correspondence between $\alpha$ clustering and the SM deformed 4p-4h quadrupole correlations was clearly established. More recently, such features were also seen in the case of $^{40}$Ca another doubly magic nucleus (see below). \\ As seen previously, $\alpha$ cluster states in light even-even nuclei with N=Z are generally connected by strong E2 transitions of several tenths of W.u. But it has also been shown that in nuclei with N different from Z, the E1 transitions between cluster states could be particularly strong. We have mentioned previously the case of $^{10}$Be where the 0$^+$ and 1$^-$ cluster states at 6.18 and 5.96 MeV are connected by a strong E1 transition of 4.3 $\pm$ 1.6 $\times$ 10$^{-2}$ W.u. The case of $^{18}$O is even more remarkable, $\alpha$-$^{14}$C cluster states with positive and negative parity have been clearly identified through the $\alpha$ transfer reaction $^{14}$C($^{7}$Li,t) and the resonant radiative capture reaction  $^{14}$C($\alpha$,$\gamma$) \cite{gai}. It has been demonstrated that the $\alpha$ cluster states with large spectroscopic factors in the transfer reactions have also enhanced E1 decay transitions. While for the `normal' $^{18}$O states with SM $^{16}$O+2n structure, the B(E1) transition strength is in the 10$^ {-3}$ to 10$^ {-4}$ W.u. range, this B(E1) strength for transitions between the $^{18}$O `pear-shaped' $\alpha$-$^{14}$C cluster states is larger than 10$^ {-2}$ W.u. \cite{gai}. It is worth mentioning that in the heavy nucleus $^{212}$Po, states populated selectively by the $\alpha$ transfer reaction $^{208}$Pb($^{18}$O,$^{14}$C) also show strong E1 decay strengths between 10$^ {-3}$ and 2 $\times$ 10$^ {-2}$ W.u. \cite{ast}. The $^{212}$Po states populated are probably of $\alpha$-$^{208}$Pb cluster structure.\\
In the last decade, one of the most remarkable discovery in nuclear structure of light nuclei has been the observation of superdeformed (SD) rotational bands in the N=Z nuclei $^{36}$Ar and $^{40}$Ca close to the end of the sd shell. These bands were fed by fusion-evaporation reactions: $^{24}$Mg($^{20}$Ne,2$\alpha$) in the case of $^{36}$Ar \cite{sven} and $^{28}$Si($^{20}$Ne,2$\alpha$) in the case of $^{40}$Ca \cite{ide}. The 0$^+$ band heads of these bands are located at E$_x$ = 4.33 MeV in $^{36}$Ar and at E$_x$ = 5.21 MeV in $^{40}$Ca. For these very deformed prolate bands, the deformation parameter $\beta _2$ is 0.46 $\pm$ 0.03 in $^{36}$Ar and 0.59 $\pm$ 0.09 in $^{40}$Ca. In the two SD bands, the B(E2) strength of the  4$^+$ $\rightarrow$ 2$^+$ transition is 53 $\pm$ 9 W.u. for $^{36}$Ar and 170 $\pm$ 40 W.u. for $^{40}$Ca. In the case of $^{40}$Ca, the SD band is nicely described by SM calculations \cite{cau} using a kind of ZBM interaction \cite{zbm} with a $^{28}$Si core and allowing up to 8 jumps of nucleons from the sd to pf shell. In $^{40}$Ca, there are three 0$^+$ states below an excitation energy of 5.5 MeV, their structure in terms of SM np-nh states is given in Table \ref{table1} \cite{cau}. 
\begin{center}
\begin{table}[h]
\caption{\label{table1} Table extracted from Ref.\cite{cau} which shows the percentage of np-nh shell model wave function components and the energy (in MeV) of the first three 0$^+$ states (GS, ND and SD) of $^{40}$Ca.}
%\footnotesize\rm
\centering
%\begin{tabular}{@{}*{7}{l}}
\begin{tabular}{llllllll}
%\br
%Option&Description\\
%\mr
\hline
 & 0p-0h & 2p-2h & 4p-4h & 6p-6h & 8p-8h & E(th) & E(exp) \\
\hline \hline
0$^+_{GS}$ & 65 & 29 & 5 & - & -& 0 & 0 \\
0$^+_{ND}$ & 1 & 1 & 64 & 25 & 9 & 3.49 & 3.35 \\
0$^+_{SD}$ & - & - & 9 & 4 & 87 & 4.80 & 5.21 \\
\hline
\end{tabular}
\end{table}
\end{center}

The $^{40}$Ca ground state 0$^+_{GS}$ is spherical with 65\% 0p-0h component in its wave function, the 0$^+$ excited state at E$_x$ = 3.35 MeV has a main 4p-4h component (64\%) and is the band head of a normal deformed (ND) band. The second 0$^+$ excited state at 5.21 MeV, is the band head of the SD band with a 87\% 8p-8h component in its wave function. In a sense, the structure of the 0$^+_{ND}$ at 3.35 MeV and of the 0$^+_{SD}$ at 5.21 MeV can be interpreted as $\alpha$-$^{36}$Ar and $^{8}$Be-$^{32}$S (or $\alpha$-$\alpha$-$^{32}$S) cluster states. These two states are indeed strongly excited in the $\alpha$ transfer reactions induced by Li beams on $^{36}$Ar targets \cite{nndc}. It is quite interesting to note here that very deformed cluster bands are observed in doubly-magic, in principle closed-shell, $^{16}$O and $^{40}$Ca nuclei. These cluster bands are of np-nh structure and built on excited and highly deformed 0$^+$ states. The `physics' involved here is very similar to what is called the island of inversion where N$\sim$20 Ne, Na, and Mg isotopes show strong deformations in their ground states. In this case, the sometimes called shell erosion is nothing else than the appearance in the ground states of strong deformations due to quadrupole np-nh correlations. It is quite possible that nuclei in the island of inversion have ground states with a cluster structure which calls for future experiments using radioactive neutron-rich beams.

\section{$^{12}$C-$^{16}$O clusters in $^{28}$Si}
Alpha clustering plays an important role in the description of the ground state and excited states of light nuclei in the p shell. For heavier nuclei, in the sd-shell, cluster configurations may be based on heavier substructures like $^{12}$C and $^{16}$O. These have been discussed to appear in $^{24}$Mg($^{12}$C-$^{12}$C) and $^{28}$Si($^{12}$C-$^{16}$O) both theoretically and experimentally \cite{ichi,kan,sand}. The case of the mid-sd shell nucleus $^{28}$Si is of particular interest as it shows the coexistence of deformed and cluster structures at rather low excitation energies. Its ground state is oblate, with a partial ${\alpha}$-$^{24}$Mg structure, two prolate normal deformed bands are found, one built on the 0$^+ _2$  at 4.98 MeV and on the 0$^+ _3$ at 6.69 MeV. Recently, a candidate superdeformed band with a pronounced $\alpha$-$^{24}$Mg structure has been suggested by Jenkins et al. \cite{jen28si}. In this band, the 2$^+$ (9.8 MeV), 4$^+$ and 6$^+$ members have been identified. The 0$^+$ bandhead remains to be discovered. We shall discuss here a resonant cluster band which is predicted to start close to the Coulomb barrier of the $^{12}$C+$^{16}$O collision, i.e. around 25 MeV excitation energy in  $^{28}$Si \cite{ichi,kan}. This band could be related to the breakup band of $^{28}$Si in $^{12}$C+$^{16}$O observed some time ago by the CHARISSA collaboration with a dedicated setup \cite{met}. This was observed in the $^{12}$C($^{20}$Ne,$^{12}$C$^{16}$O)${\alpha}$ reaction up to 50 MeV excitation energy. Spin systematics  has been deduced from the corresponding angular correlation measurement and the breakup states are grouped in a band which moment of inertia is close to a very deformed, molecular like $^{28}$Si. Breakup states with low spins ($\le$ 6) corresponding to low excitation energy have not been measured in this study due to experimental limitations. Our experimental work aimed at measuring resonant states in this low spin region, possibly the low-lying members of this breakup band. We have studied the $^{12}$C($^{16}$O,$\gamma$)$^{28}$Si radiative capture reaction at 5 resonant energies around the Coulomb barrier (CB). The $^{12}$C+$^{16}$O CB energy is $\sim$ 7.9 MeV \cite{lebh}. This reaction was already investigated in the 1980ies close to the CB in an experimental campaign at Brookhaven National laboratory by Sandorfi et al. \cite{sand}. Narrow resonances were identified and significant radiative capture cross-sections to some low-lying members of the $^{28}$Si ground state band were measured making use of a large volume NaI detector coupled, in one of their experiments, to a Wien velocity filter for better recoil selection \cite{col}. 
Unlike the Sandorfi measurement \cite{sand}, our data do not show the piling-up of low energy transitions due to fusion-evaporation events, and thus our spectra extends to lower energies. This may have led Sandorfi to miss some flux mediated via low-energy transitions in the decay of the resonances and thus to miss some part of the capture cross-section. 

We have decided to measure the full electromagnetic decay of some resonances in the energy region explored by Sandorfi et al., making use of the state-of-the-art recoil 0$^\circ$ spectrometer Dragon installed at Triumf, Vancouver, to identify the $^{28}$Si recoils with an incident beam rejection factor better than 10$^{12}$ and the associated 30 BGO array for the coincident high efficiency $\gamma$-ray detection. Details about the setup can be found in \cite{hutch}. For our two $^{12}$C+$^{16}$O radiative capture studies, five $^{16}$O bombarding energies have been used between 21.1 and 15.4 MeV, three above the CB, on- and off-resonances observed by Sandorfi, and two below CB: the first on a previously observed resonant structure and the lower one below all previous radiative capture measurements was chosen on the maximum of a structure observed in the fusion excitation function \cite{chris}. A very efficient selection of the radiative capture channel out of the five orders of magnitude dominant fusion-evaporation events was performed by setting a gate on the time-of-flight between the DSSSD detector at the Dragon focal plane and signals of the 21m upstream BGO array.\\
Monte Carlo GEANT calculations have been performed to simulate the response function of the experimental setup to different entrance channel conditions, i.e. energy and spin of the resonance.
%T¨hese calculations allowed to take into account the full  experimental setup, to track down the $^{28}$Si %recoils through the Dragon spectrometer and to calculate the corresponding $\gamma$ spectrum. More %details about the procedure can be found in \cite{lebh}, which concerns our radiative capture studies at %energies above the CB.
\begin{figure}[h]
\begin{center}
\includegraphics[width=12cm]{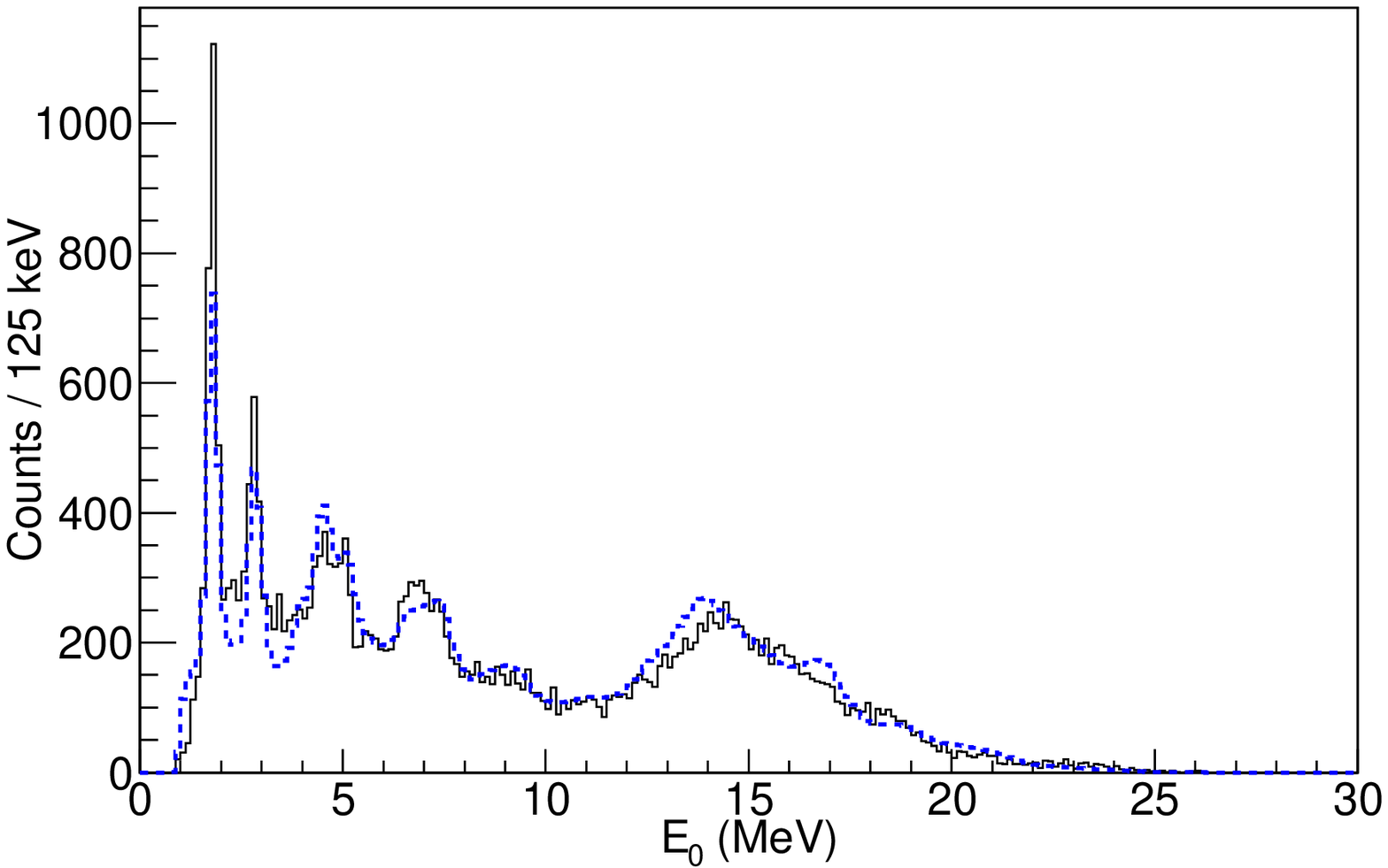}
\includegraphics[width=12cm]{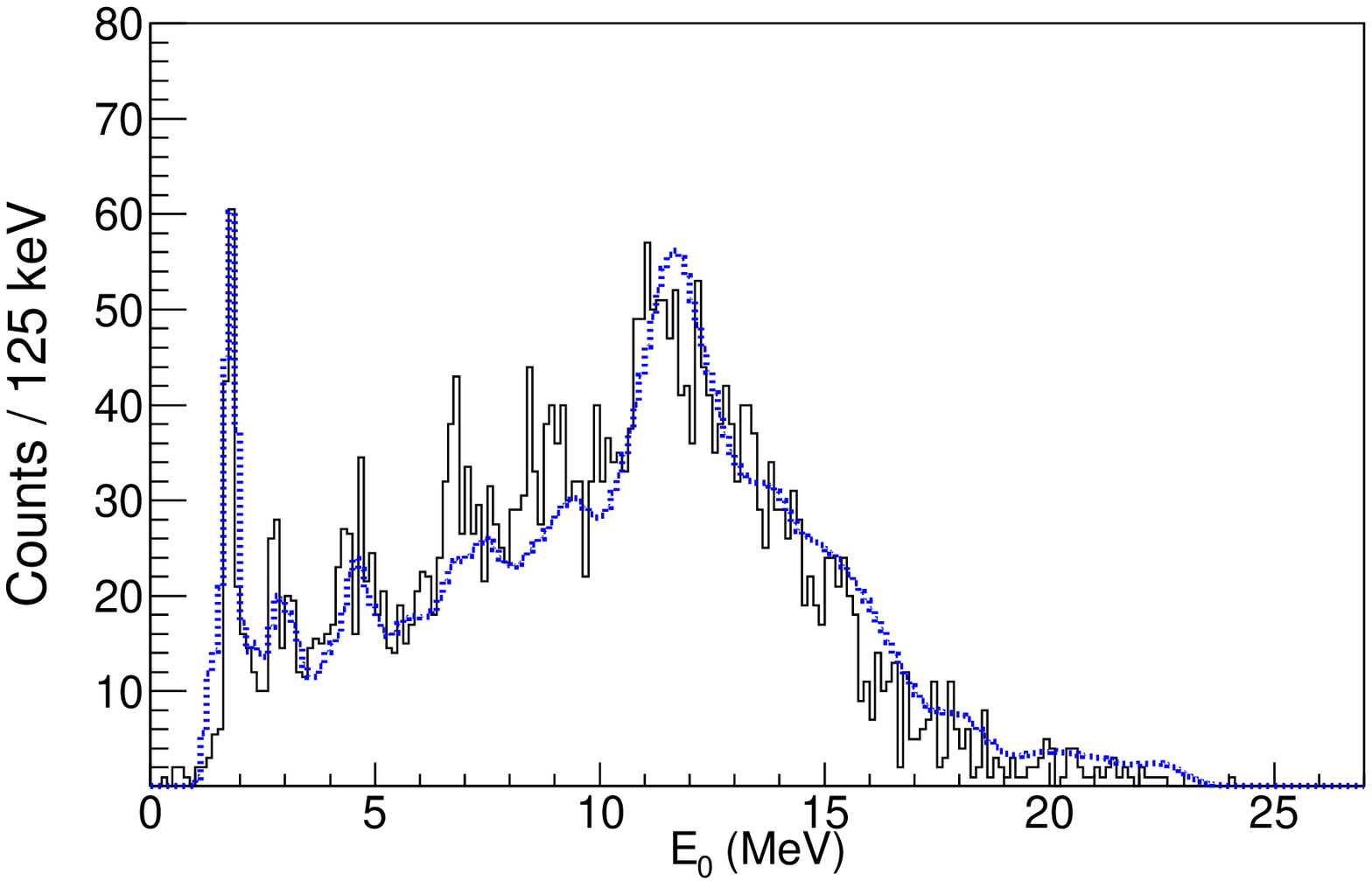}
\end{center}
\caption{\label{fig1} (Color online) Gamma spectra of the highest energy $\gamma$-ray in the cascade (E$_0$) at E$_{c.m.}$ = 9 MeV (top) and at E$_{c.m.}$ = 6.6 MeV (bottom) and Monte-Carlo simulations (blue dashed line) of the decay spectrum for a 6$^+$ entrance resonance (top) and for a 2$^+$ entrance resonance (bottom). }
\end{figure}

\begin{figure}[h]
\begin{center}
\includegraphics[width=5cm]{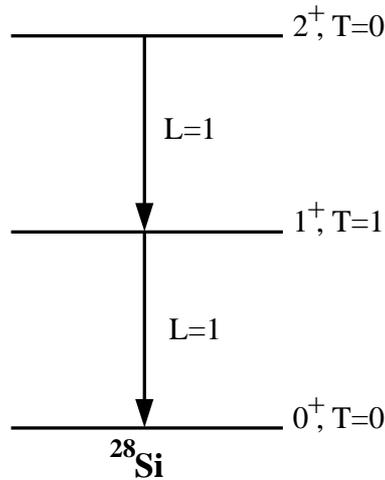}
\end{center}
\caption{\label{fig3} Schematic view of the decay scenario of the lowest energy 2$^+$ resonance (E$_{c.m.}$ = 6.6 MeV / E$^*$($^{28}$Si) = 23.4 MeV) via isovector M1 transitions feeding an intermediate T=1 state.}
\end{figure}

Figure \ref{fig1} shows the $\gamma$ spectra of the highest energy $\gamma$-ray (E$_0$) in the cascade measured in the BGO array at two bombarding energies of our study, the highest energy E$_{Lab}$ = 21.1 MeV (E$_{c.m.}$ = 9 MeV) at the top and the lowest bombarding energy E$_{Lab}$ = 15.4 MeV (E$_{c.m.}$ = 6.6 MeV) at the bottom. Below 10 MeV, the spectra show $\gamma$ transitions between $^{28}$Si low lying states and above an unexpected bump around E$_0$ = 14 and 12 MeV. This bump results from the feeding of intermediate states in the decay in an energy region above the $\alpha$ emission threshold (at 9.98 MeV). Due to the BGO array resolution, individual peaks are not resolved in this bump. Data is compared to numerical simulations (blue dashed curves on Fig. \ref{fig1}). These simulation take into account the entire experimental setup, i.e. the BGO array, the Dragon 0$^{\circ}$ spectrometer optical elements and focal plane detection. The $^{28}$Si recoils are fully tracked down through the Dragon spectrometer. Characteristics of the $^{28}$Si compound nucleus are used in the simulations, such as entrance excitation energy, bound and quasi-bound states from the literature and their $\gamma$ branching ratios \cite{nndc}. The entrance angular momentum is the main parameter of the simulations and the $\gamma$ spectra obtained take into account the electromagnetic transition properties in the $^{28}$Si mass region \cite{endt}. The best fit of the data is obtained for :
\begin{itemize}
\item A 6$^+$ entrance resonance at the highest bombarding energy (E$_{Lab}$ = 21.1 MeV / E$_{c.m.}$ = 9.0 MeV). The data clearly show the direct feeding of the prolate deformed 4$^+_3$ state at 9.16 MeV and of the octupole deformed 3$^-$ at 6.88 MeV. This state is the band head of an octupole band which mainly decays to the $^{28}$Si oblate ground state with a strong E3 transition of reduced width B(E3) = 20 W.u. Our results are very similar to what has been measured for the $^{12}$C+$^{12}$C radiative capture reaction above the CB in a previous Triumf Dragon experiment \cite{jenk} where the enhanced feeding of the $^{24}$Mg prolate band has been measured from a 4$^+$-2$^+$ resonance at E$_{c.m.}$ = 8.0 MeV near the CB.
\item A 2$^+$ entrance resonance at the lowest bombarding energy (E$_{Lab}$ = 15.4 MeV / E$_{c.m.}$ = 6.6 MeV). Our data clearly show the enhanced feeding, directly from the resonance, of J$^{\pi}$=2$^+$ and 1$^+$ T=1 states around 11 MeV. The main flux is going to the 1$^+$ state at 11.45 MeV which decays mainly to the $^{28}$Si ground state by an isovector M1 transition. This scenario is confirmed by experimental angular distributions which show that dipole transitions are emitted from the 2$^+$ entrance resonance, which feed the intermediate T=1 states around 11 MeV, which then also decay via dipole transitions to the $^{28}$Si ground state (see simplified view of this scenario on Fig. \ref{fig3}). At this energy, the decay is again quite similar to what occurs in the $^{12}$C+$^{12}$C system at the lowest explored energy of Ref. \cite{jenk}. Here as well,  J$^{\pi}$=2$^+$ has been assigned to the entrance resonance and an enhanced decay has been measured via intermediate 1$^+$ T=1 states around 11 MeV in $^{24}$Mg. A definitive scenario  for the decay of the resonances at these low bombarding energies in both systems would come from the measurement of the $\gamma$ decay spectra with a $\gamma$ spectrometer with better resolution than BGO but still rather good efficiency. This type of array, based on LaBr$_3$ crystals is under construction by the PARIS collaboration \cite{maj} and should be available in the near future for such radiative capture studies.
\end{itemize}
Moreover, it would be interesting to determine if the decay scenario involving M1 $\Delta$T=1 transitions is specific to self conjugate nuclei. This could be investigated by performing for instance the $^{12}$C+$^{14}$C radiative capture reaction.
%\end{itemize}

Interestingly enough, our spin assignments around the CB agree very well with the previously mentioned breakup band spin systematics measured by the CHARISSA collaboration and our resonant states could be the low lying members of the very deformed molecular-like $^{12}$C-$^{16}$O band they have identified \cite{met}. \\
What higher bombarding energies are concerned, it would be interesting to search for the evolution of the observed feeding of the $^{28}$Si intermediate states in the decay of the resonances. A pioneering campaign of experiments has been performed by Harakeh et al. measuring the isospin mixing in the $^{28}$Si excitation energy region around 35 MeV via the $^{12}$C($^{16}$O,$\gamma$)$^{28}$Si radiative capture reaction \cite{har}. The authors discuss their $\gamma$ spectra in terms of the feeding of the giant dipole resonance built on $^{28}$Si excited states. It could be possible that in our studies of the electromagnetic decay of $^{12}$C+$^{16}$O resonances near CB, the giant quadrupole resonance (GQR) is excited, built on the $^{28}$Si prolate band starting at $\sim$7 MeV. Such a mechanism would thus populate states located $\sim$7 MeV above the maximum of the GQR,  measured at 18.8 MeV \cite{young}, i.e. a region around 26 MeV, close to our resonance excitation energies.

\end{document}